\documentclass[runningheads,citeauthoryear]{apinv}
\usepackage{epsfig,cite,graphics}
\usepackage[T2A]{fontenc}
\usepackage[cp1251]{inputenc}

\begin{document}

\title{First observations with the 25 cm telescope of the Shumen Astronomical Observatory}
\titlerunning{First observations with the 25 cm telescope of ShAO}
\author{Diana Kjurkchieva\inst{1}, Sunay Ibryamov\inst{1}, Borislav Borisov\inst{1}, Dragomir Marchev\inst{1}, Velimir Popov\inst{1}, Dinko Dimitrov\inst{2,1}}
\authorrunning{D. Kjurkchieva, S. Ibryamov, B. Borisov, D. Marchev, V. Popov, D. Dimitrov}
\tocauthor{D. Kjurkchieva}
\institute{Department of Physics and Astronomy, Shumen University,
115 Universitetska, 9700 Shumen, Bulgaria \and Institute of
Astronomy and NAO, Bulgarian Academy of Sciences, Tsarigradsko
shossee 72, 1784 Sofia, Bulgaria
 \newline
    \email{d.kyurkchieva@shu.bg}
 }
\papertype{Submitted on xx.xx.xxxx; Accepted on xx.xx.xxxx}
\maketitle

\begin{abstract}
The first observations with the 25 cm telescope of the Shumen
Astronomical Observatory led to the following conclusions: (a) Intra-night
observations of variable stars with an amplitude larger than 0.1 mag
are possible down to 14 mag with an acceptable quality with this setup;
(b) The equipment is suitable for observations of bright
extended objects with sizes up to 30 arcmin (planets, comets,
clusters, nebulae, galaxies) with resolution 0.88 arcsec/pix; (c)
The guiding of telescope is very good which makes the equipment
appropriate for prolonged patrols; (d) The observations with
the 25 cm are already fully remote-controlled; (e) The determined
transformation coefficients allow transfer from instrumental to
standard photometric system $BVR_cI_c$ and realization of
differential photometry.
\end{abstract}


\section*{Introduction}

The invention of remote controlled telescopes at the end of the
20-th century increased the role of small telescopes in the
observational astronomy and efficiency of usage of observational
time (Iliev 2014).

The small telescopes have an invaluable role in the discovery of
variable stars of different type by ground-based wide-field
surveys: ASAS (Pojmanski 1997), ROTSE (Akerlof et al. 2005),
SuperWASP (Pollacco et al. 2006), CRTS (Drake et al. 2014), etc.
The detailed investigation of these stars is also appropriate task
for small telescopes. The main goal of this paper is to
demonstrate the possibilities of the 25 cm telescope of Shumen
Astronomical Observatory (ShAO) for observations of objects of
different type, mainly variable stars.

\section*{1. Equipment}

The 25 cm Schmidt-Cassegrain telescope \emph{MEADE 10$''$ LX80 SC}
(focal length 2540 mm) is driven by mounting \emph{Sky-Watcher
NEQ6 PRO} fixed on a special column. It is designed to provide not
only stability but also free movement of the telescope in all
directions. The optical equipment is situated inside a 3
m automated telescope dome (Fig. 1) in the ShAO, whose
coordinates are: longitude 26$^{\circ}$55$'$23$''$ E, latitude
+43$^{\circ}$15$'$27$''$ N, altitude 493 m.

\begin{figure}[!htb]
  \begin{center}
    \centering{\epsfig{file=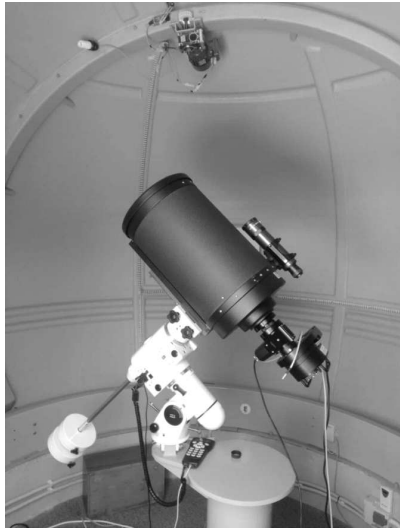, width=0.4\textwidth}}
    \caption[]{The 25 cm telescope inside the 3 m dome}
    \label{fig1}
  \end{center}
\end{figure}

The telescope is equipped with a detector \emph{SBIG
ST-10XME} (2184 $\times$ 1472 pixels, 6.8$\mu$m/pixel) and
a SBIG CFW10 filter wheel with circle 1.25-inch Johnson-Cousins
mounted filters $BVR_cI_c$. This telescope-detector system
provided field of view (FoV) of 20 $\times$ 14 arcmin with
resolution 0.55 arcsec/pix. The FoV was increased by focal reducer
\emph{TS Optics} f/6.3 to 32 $\times$ 22 arcmin with resolution
0.88 arcsec/pix. The remote focusing is performed by focuser
\emph{Optec Temperature Compensating Focuser}.

Generator \emph{ProTech P12000} and \emph{ATG} system (Fig. 2) can
provide electricity supply of the equipment during 5 hours in case
of accidental events.

\begin{figure}[!htb]
  \begin{center}
    \centering{\epsfig{file=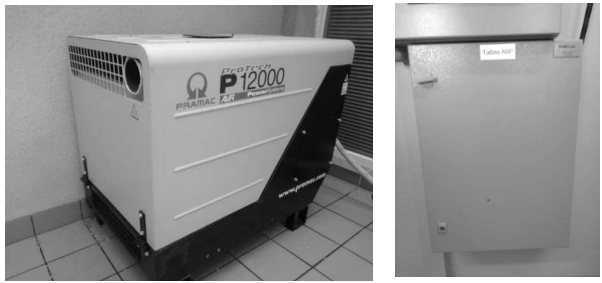, width=0.6\textwidth}}
    \caption[]{Generator \emph{ProTech P12000} and \emph{ATG} system}
    \label{fig2}
  \end{center}
\end{figure}

\begin{figure}[!htb]
  \begin{center}
    \centering{\epsfig{file=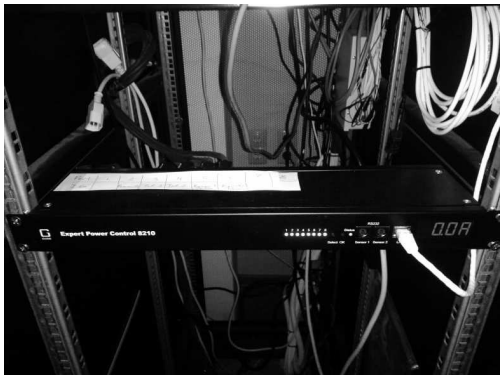, width=0.5\textwidth}}
    \caption[]{Computer controlled power switch \emph{Expert Power Control 8210}}
    \label{fig3}
  \end{center}
\end{figure}

The components of the whole equipment were connected to a computer
controlled power switch \emph{Expert Power Control 8210} (Fig. 3).
A remote-control videocamera inside the 3 m dome allows us to see
the telescope and shutter positions.

Cloud detector \emph{AAG CloudWatcher} provides full
information of sky conditions: (i) Its light sensor allows to
estimate the sky glow; (ii) An infrared sensor measures the
outside temperature while an electronic thermometer measures the
internal temperature. Both data determine the existence of clouds;
(iii) A variable capacitor determines the existence of rain.

\section*{2. Software and remote control}

The remote-control observations by the 25 cm telescope require the
following software: (i) \emph{C2A} realizes telescope remote
control; (ii) \emph{FocusMax V4} allows remote focusing; (iii)
\emph{MaxIm DL Version 6} controls the CCD camera, filter wheel
and focuser: (iv) \emph{CCDAutoPilot 5.0 Professional} connects
and manages the software and drivers in a single complex; (v)
\emph{ScopeDome} drives the 3 m dome.

The \emph{Sky-Watcher NEQ6 PRO} mounting is designed for manual
control of the telescope by a special console. Such a mode
requires an alignment of the telescope every night before
observations by 2-3 stars. Our mounting does not accept time
information from a computer but only from the console that caused
a huge problem for the realization of remote-control observations.
After many tests, we managed to overcome the difficulties by using
\emph{Astrometry.net}. Currently the 25 cm telescope is
remote-controlled and further we present results of the first
observations.

\section*{3. Images of extended objects}

The investigation of extended objects requires adequate background
estimation. The image reduction techniques aim to diminish the
impact of both instrumental and non-instrumental offsets, which
may lead to erroneous flux estimation (Popowicz $\&$ Smolka 2015).
The instrumental sources of background variations are related to
the CCD detectors and the most of them are mitigated recently by
novel CCD structures, proper CCD calibration and strong cooling.

The main source of non-instrumental background variations is the
sky glow, which is dependent on the altitude above the horizon and
the observed wavelength range. It originates from the local
light pollution and is time-dependent. The contribution of the
sky glow is bigger in the infrared. There are techniques to
subtract the background bias frames in real-time (Bertero et al.
2000; Fiorucci et al. 2003) but the scattered light in the
instrument from nearby very bright object is still difficult to
remove.

Figures 4--8 illustrate the image quality of extended objects
(planets, comets, nebulae, etc.) observed by our 25 cm telescope
and different detectors. They reveal that the city lights have not
noticeable negative contribution to the image quality.
Surprisingly, the lights of the moving cars alongside the
observatory also do not hinder the image quality considerably.
Hence, the 25 cm telescope might be used for study of extended
objects with size up to 30 arcmin (planets, comets, clusters,
nebulae, galaxies) with resolution 0.88 arcsec/pix. Images of
objects with a larger angular size might be obtained by mosaic observations.

\begin{figure}[!htb]
  \begin{center}
    \centering{\epsfig{file=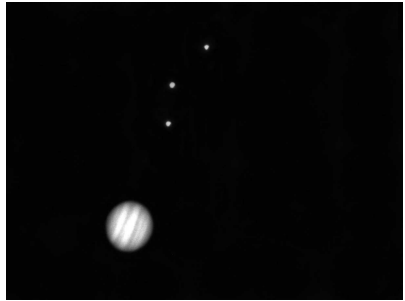, width=0.4\textwidth}}
    \caption[]{Jupiter with three Galilean satellites (Io, Ganymede and Europa), exposure 1 sec, 2017 April 10, detector \emph{NexImage 5}}
    \label{fig4}
  \end{center}
\end{figure}

\begin{figure}[!htb]
  \begin{center}
    \centering{\epsfig{file=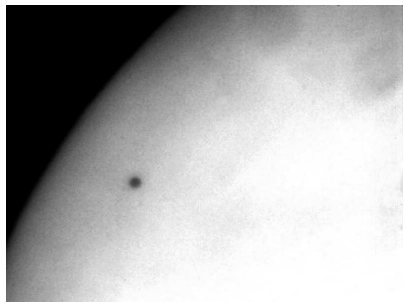, width=0.4\textwidth}}
    \caption[]{Transit of Mercury, 2016 May 9, detector \emph{NexImage 5}}
    \label{fig5}
  \end{center}
\end{figure}

\begin{figure}[!htb]
  \begin{center}
    \centering{\epsfig{file=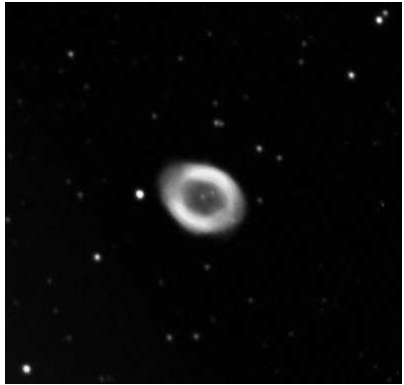, width=0.4\textwidth}}
    \caption[]{Planetary nebula M57, 2018 Aug 3, detector \emph{SBIG ST-10XME}, exposure 180 s, R filter}
    \label{fig6}
  \end{center}
\end{figure}

\begin{figure}[!htb]
  \begin{center}
    \centering{\epsfig{file=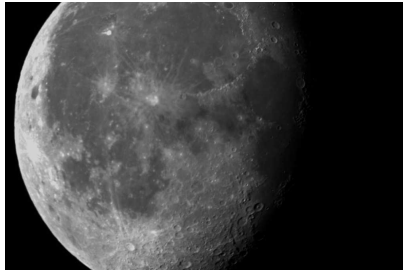, width=0.4\textwidth}}
    \caption[]{The Moon, 2018 Aug 30, detector \emph{SBIG ST-10XME}, exposure 0.1 s, B filter}
    \label{fig7}
  \end{center}
\end{figure}

\begin{figure}[!htb]
  \begin{center}
    \centering{\epsfig{file=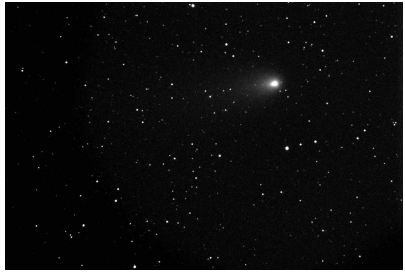, width=0.4\textwidth}}
    \caption[]{The comet 21P/Giacobini-Zinner, 2018 Aug 30, detector \emph{SBIG ST-10XME}, exposure 30 s, R filter}
    \label{fig8}
  \end{center}
\end{figure}

\section*{4. Investigation of stellar variability}

We chose to test the possibilities of the 25 cm telescope and its
equipment for study of stellar variability by observations of the
W UMa star V568 Peg. The reasons for this choice were: (i) The
target has been observed earlier also by small telescope
(Kjurkchieva et al. 2015) which gives an opportunity for comparison
of the results; (ii) The target V magnitude is 13.5 which was
the expected highest value, at which our photometric precision
would be acceptable; (iii) The short-period of around 6 hrs of
the binary was important to obtain observations covering the whole
cycle during a night taking into account the very bad atmospheric
conditions in the last seasons of 2018.

\begin{figure}[!htb]
  \begin{center}
    \centering{\epsfig{file=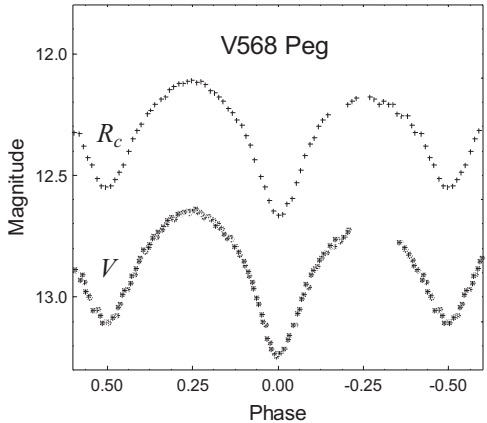, width=0.5\textwidth}}
    \caption[]{The folded $V,R_c$ light curves of V568 Peg from 2018 Aug 13 obtained by the 25 cm telescope and CCD detector \emph{SBIG ST-10XME}}
    \label{fig9}
  \end{center}
\end{figure}

V568 Peg was observed on 13 Aug 2018. The atmospheric conditions
at the beginning of the night (UT = 19 h) were good: clear sky,
temperature 25 C, humidity 50 $\%$. The CCD was cooled to -10 C.
There was no wind and Moon (night after new Moon). The exposures
in V and R filter were respectively 90 s and 60 s. The initial
FWHM of the target images were around 4 pix in V and 5.5 pix in R.
After the midnight the target images became considerably better
with FWHM up to 2.4 pix in V and 4.5 pix in R. Unfortunately 2 hrs
later the outside temperature decreased and humidity rapidly
increased. The FWHM of the target images did not change
considerably but the signal rapidly decreased. We stopped the
observations when the humidity reached 90 $\%$.

It should be pointed out that there was not any need to make
corrections in the telescope guiding during the 7-hr observations.

Standard procedure was used for reduction of the photometric data.
We performed differential photometry using nearby standard stars.
The obtained folded $V,R_c$ light curves of V568 Peg are shown in
Fig. 9. The average photometric accuracy is 0.029 in both filters.

\begin{figure}[!htb]
  \begin{center}
    \centering{\epsfig{file=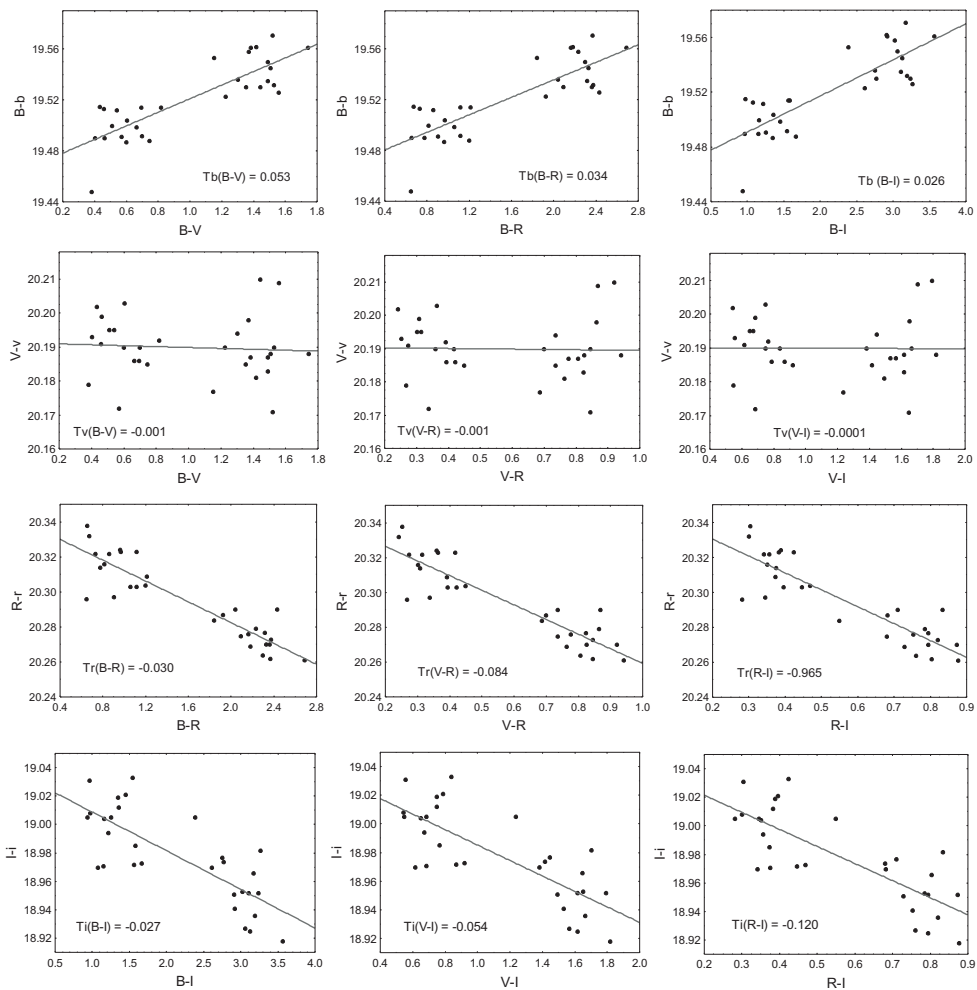, width=0.99\textwidth}}
    \caption[]{Diagrams for determination of the first set of transformation coefficients}
    \label{fig10}
  \end{center}
\end{figure}

\begin{figure}[!htb]
  \begin{center}
    \centering{\epsfig{file=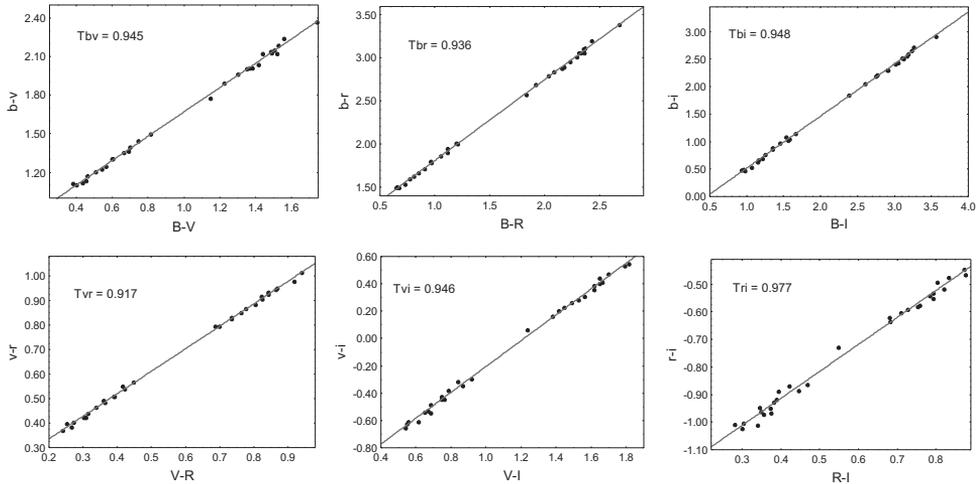, width=0.99\textwidth}}
    \caption[]{Diagrams for determination of the second set of transformation coefficients}
    \label{fig11}
  \end{center}
\end{figure}

\begin{table}[tp]\footnotesize
\begin{center}
\caption[]{Transformation coefficients
 \label{t1}}
 \begin{tabular}{clclcl}
\hline\hline
coefficient&value& coefficient &  value & coefficient &  value   \\
  \hline
Tbv   & 0.945 & Tb(B-V) & 0.053  & Tr(B-R) & -0.030  \\
Tbr   & 0.936 & Tb(B-R) & 0.034  & Tr(V-R) & -0.084  \\
Tbi   & 0.948 & Tb(B-I) & 0.026  & Tr(R-I) & -0.965  \\
Tvr   & 0.917 & Tv(B-V) &-0.001  & Ti(B-I) & -0.027  \\
Tvi   & 0.946 & Tv(V-R) &-0.001  & Ti(V-I) & -0.054  \\
Tri   & 0.977 & Tv(V-I) &-0.0001 & Ti(R-I) & -0.120  \\
  \hline
\end{tabular}
\end{center}
\end{table}

\section*{5. Transformation coefficients}

The measurements of star brightness need to be transformed from
instrumental to standard photometric systems. For this purpose, we
followed the procedure given in the CCD Photometry Guide of the
American Association of Variable Star Observers
(https://www.aavso.org/ccd-photometry-guide).

The determination of the transformation coefficients requires
observing as many as possible standard stars with a wide range of
colors. In order to obtain the transformation coefficients of our
equipment we carried out observations of the open cluster NGC 7790
in $BVR_cI_c$ filters with exposures of 60 s on 27 Sept 2018. The
atmospheric conditions were excellent. We measured 30 stars in the
field and the results are shown in Figs. 10--11 and Table 1.

The transformation coefficients allow to perform differential
photometry if the calibrated magnitudes of suitable comparison
stars in the field have been previously determined. Then only the
color and magnitude differences between the target and the
comparison stars are important because all they have the same
airmass.

Future observations will allow to assess the stability of the
transformation coefficients of our equipment on long time scales.

\section*{Conclusions}

The first observations with the 25 cm telescope of Shumen
Astronomical Observatory led to the following conclusions.

(1) The equipment is suitable for observations of bright
extended objects with sizes up to 30 arcmin (planets, comets,
clusters, nebulae, galaxies) with resolution 0.88 arcsec/pix.

(2) Intra-night observations of variable stars with an
amplitude larger than 0.1 mag are possible down to 14 mag with an
acceptable quality with this setup.

(3) The guiding of telescope is very good which makes the
equipment appropriate for prolonged patrols.

(4) The sky glow doesn't have any noticeable negative
contribution to the image quality.

(5) The observations with the 25 cm are already fully
remote-controlled.

The determined transformation coefficients allow transfer from
instrumental to standard photometric system $BVR_cI_c$ and
realization of differential photometry.

\section*{Acknowledgments}

The research was supported partly by projects DN08-20/2016,
DN08-01/2016 and DM08-02/2016 of Scientific Foundation of the
Bulgarian Ministry of Education and Science, project
D01-157/28.08.2018 of the Bulgarian Ministry of Education and
Science as well as by projects RD-08-142/2018 and RD-08-112/2018
of Shumen University.

The authors are very grateful to the anonymous Referee for
the valuable notes and recommendations.

It used the SIMBAD database, operated at CDS, Strasbourg, France,
USNO-B1.0 catalogue (http://www.nofs.navy.mil/data/fchpix/), and
NASA's Astrophysics Data System Abstract Service.



\begin{thebibliography}{}

\bibitem{}
Akerlof C., 2005, MPC, 54971, 10

\bibitem{}
Bertero, M., Boccacci, P., Robberto, M., 2000, PASP, 112, 1121

\bibitem{}
Drake, A. J., Djorgovski, S. G., Garcia-Alvarez, D., Graham, M. J., Catelan, M. et al., 2014, ApJ, 790, 157

\bibitem{}
Fiorucci, M., Persi, P., Busso, M., Ciprini, S., Corcione, L., Tosti, G., 2003, MSAIS, 2, 125

\bibitem{}
Iliev, I., 2014, CoSka, 43, 169

\bibitem{}
Kjurkchieva, D., Popov, V., Petrov, N., Ivanov, E., 2015, CoSka, 45, 28

\bibitem{}
Pojmanski G., 1997, AcA, 47, 467

\bibitem{}
Pollacco, D. L., Skillen, I., Collier Cameron, A., Christian, D. J., Hellier, C. et al., 2006, PASP, 118, 1407

\bibitem{}
Popowicz A., Smolka B., 2015, MNRAS, 452, 809

\end{thebibliography}
\end{document}